\documentstyle{article}

\begin{document}

\pagestyle{empty}

\title{Large Lepton Mixing in Seesaw Models\thanks{
Talk is given by T. Yanagida}\\
- Coset-space Family Unification -}
\author{J. Sato and T. Yanagida\\
Department of Physics and RESCUE, University of Tokyo,\\
Hongo, Bunkyo-ku, Tokyo, 113, Japan
}

\date{\today}

\maketitle

\begin{abstract}
We show that the large mixing between $\nu_\mu$ and $\nu_\tau$
observed by the SuperKamiokande collaboration is a quite natural
prediction in a large class of seesaw models. This large mixing is basically
due to the unparallel family structure suggested from the observed mass
hierarchies in quark and lepton mass matrices.
We show that the unparallel family structure is automatically realized
in ``coset-space family unification''
model based on E$_7/$SU(5)$\times$U(1)$^3$.
This model also suggests the small angle MSW solution to the solar
neutrino problem.
\end{abstract}

\section{Introduction}

T. Kajita from the SuperKamiokande collaboration has reported, in this
conference, very convincing evidence of neutrino oscillation in their
atmospheric neutrino data\cite{SK}. It is now clear that the long-standing
puzzle of muon neutrino deficit in underground detectors\cite{Atm}
is due to the neutrino oscillation. A remarkable feature of the oscillation
is almost maximal mixing between $\nu_\mu$ and $\nu_\tau$ ($\sin^2\theta_{23}
\ge 0.8$), in sharp contrast to the quark sector for which mixing angles among
different generations are all small. At first glance the rule governs
the lepton mass matrices seems significantly different from the one
relevant for the quark sector. We first show, in this talk, that the
large mixing between $\nu_\mu$ and $\nu_\tau$ is quite naturally understood
in a large class of seesaw models\cite{seesaw}.

\section{General Consideration in Seesaw Models}

We adopt the SU(5) grand unification (GUT) as an example to make our
point clearer, in which the lepton doublets belong to ${\bf 5^*}$ of SU(5)
GUT. We also assume supersymmetry(SUSY).

Let us discuss first the up-type quark mass matrix that is given by
the following superpotential:

\begin{equation}
W= h_{ij}{\bf 10}_i {\bf 10}_j <H({\bf5})>.
\end{equation}

The most natural explanation of the   mass hierarchy is given by the
Froggatt-Nielsen mechanism\cite{FN}. We here assume a U(1) symmetry
which is broken by a condensation of a superfield $\phi$.
The observed mass hierarchy, 
\begin{equation}
m_t : m_c : m_u \simeq 1 : \epsilon^2 : \epsilon^4,
\end{equation}
suggests that $\epsilon=<\phi>/M_G \sim 1/20$
and the U(1) charges are 0, 1, 2
and -1 for the third, second, first families of {\bf 10}'s and the $\phi$.
Here $M_G$ is the gravitational scale $M_G\simeq2.4\times 10^{18}$GeV.

The down-type quark/charged lepton mass matrix is given by
\begin{equation}
W=f_{ij}{\bf 10}_i {\bf 5}^*_j <\overline{H}({\bf 5}^*)>.
\end{equation}
The observed mass hierarchy,
\begin{equation}
m_b:m_s:m_d = m_\tau:m_\mu:m_e\simeq 1:\epsilon:\epsilon^3,
\end{equation}
suggests that the third, second and first families of ${\bf 5}^*$ have the
U(1) charges $A$, $A$, and $A+1$, respectively. A crucial point is that
the third and the second families of ${\bf 5}^*$ have the same U(1) charge
$A$. $A$ could be 0 or 1. We take $A=0$ for simplicity. We should
stress here that the observed mass hierarchies in quark and lepton mass
matrices already suggest an unparallel family structure in Table \ref{uptable}.
\begin{table}[hbt]
\begin{tabular}{ccccc}
&&&&U(1) charge\\
\hline
${\bf 10}_3$ &${\bf 5}^*_3$ & ${\bf 5}^*_2$&&[0]\\
${\bf 10}_2$ &${\bf 5}^*_1$ & &&[1]\\
${\bf 10}_1$ & & &&[2]\\
\end{tabular}
\caption{Unparallel Family Structure}
\label{uptable}
\end{table}

Now, let us discuss the neutrino mass matrix. In a generic seesaw model
it is given by the following effective superpotential:
\begin{equation}
W_{eff}={\kappa_{ij}\over M_{\nu_R}} {\bf 5}_i^* {\bf 5}_j^* <H({\bf 5})
 H({\bf 5})>
\label{nlmass}
\end{equation}
The U(1) charge assignment for ${\bf 5}^*_i$ leads to
\begin{equation}
\kappa_{ij}\sim \pmatrix{1&1&\epsilon\cr
                          1 &1& \epsilon\cr
                            \epsilon &\epsilon &\epsilon^2}.
\label{kappa}
\end{equation}
Notice that the U(1) charges for the superheavy right-handed neutrino
$\nu_R$ are canceled out in the effective neutrino mass matrix in
eq.(\ref{nlmass}). From eq.(\ref{kappa}) we easily see a large mixing 
close to the maximal between $\nu_\mu$ and $\nu_\tau$.
The appearance of the large mixing is originated from the unparallel
family structure discussed above.\footnote{This crucial point is
emphasized by T. Yanagida and P. Ramond in this conference.}
On the contrary to the
$\nu_\mu$-$\nu_\tau$ mixing, we have small mixing between $\nu_e$ and
$\nu_\mu$ or $\nu_\tau$. Thus, the small angle MSW solution \cite{MSW}
to the solar neutrino problem \cite{Solar}
is also a quite natural expectation in 
a large class of seesaw models.

\section{Coset-space Family Unification on E$_7$/SU(5)$\times$U(1)$^3$}

In this section we show that the unparallel family structure discussed
in the previous section is naturally obtained in the coset-space family
unification\cite{E71} based on E$_7$.

The  E$_7/$SU(5)$\times$U(1)$^3$ model\cite{E72,YY} contains
three families of ${\bf 10}_i + {\bf 5}^*_i + {\bf 1}_i$
($i=1-3$)
and one ${\bf 5}$ as NG multiplets. Here, the SU(5) is the usual
GUT gauge group.
Their quantum numbers under the unbroken subgroup are given in Table
\ref{charge}. Notice that the first family ${\bf 10_1}$ has non-vanishing
charge only for the U(1)$_3$ which means that the ${\bf 10_1}$ is the NG
multiplet for SO(10)/SU(5)$\times$U(1). Similarly, we find that ${\bf 10_2}$,
${\bf 5^*_1}$ and ${\bf 1_1}$ are NG multiplets for E$_6$/SO(10)$\times$U(1)
and the remaining fields are NG multiplets for E$_7$/E$_6\times$U(1).
Thus, it is now clear that the unparallel family structure is an 
automatic prediction of this coset-space family unification \cite{SY}.

\begin{table}[tb]
\begin{center}
\begin{tabular}{c|ccc}
SU(5)
&\makebox[1.5cm]{U(1)$_1$}&\makebox[1.5cm]
{U(1)$_2$}&\makebox[1.5cm]{U(1)$_3$}\\
\hline
{\bf 10}$_1$&0&0&4\\
{\bf 10}$_2$&0&3&-1\\
{\bf 10}$_3$&2&-1&-1\\
{\bf 5}$^*_1$&0&3&3\\
{\bf 5}$^*_2$&2&-1&3\\
{\bf 5}$^*_3$&2&2&-2\\
{\bf 1}$_1$&0&3&-5\\
{\bf 1}$_2$&2&-1&-5\\
{\bf 1}$_3$&2&-4&0\\
\hline
{\bf 5}&2&2&2
\end{tabular}
\caption{U(1) charges of the NG multiplets. The U(1)$_1$,
U(1)$_2$ and U(1)$_3$ are the unbroken U(1)'s of
coset-subspaces E$_7$/E$_6\times$U(1),
E$_6$/SO(10)$\times$U(1) and SO(10)/SU(5)$\times$U(1), respectively.}
\label{charge}
\end{center}
\end{table}

This model can not be quantized in the original form,
since there is a nonlinear-sigma model anomaly\cite{Anom,YY}.
However, this global obstruction is easily removed\cite{YY}
by introducing a matter multiplet ${\bf 5}^*$
which is also needed for an SU(5) gauge-anomaly
cancellation\cite{E72}. We assume that some explicit breaking induces an
invariant mass for the NG ${\bf 5}$ and this matter ${\bf 5}^*$
and we neglect them in our discussion.

In addition to the NG multiplets we introduce a pair of Higgs
multiplets ${\bf 5}_H$ and ${\bf 5}^*_H$. As long as
the global E$_7$
is exact these Higgs multiplets never
have Yukawa couplings to the NG quarks and leptons.
Thus, the observed hierarchy in quark-lepton mass
matrices is regarded as a consequence of a hierarchy 
in the explicit breaking of the global E$_7$.
This situation is very similar to that in the QCD,
where the mass hierarchy between NG pions and kaons
($m_K^2\gg m_\pi^2$) is originated from the hierarchy
in quark masses ($m_s\gg m_{u,d}$) which are explicit
breaking parameters of the chiral SU(3)$_L\times$SU(3)$_R$.

We consider three steps for the explicit breaking:
\begin{equation}
\begin{tabular}{ccccccc}
$ {\rm E}_7$ &$\longrightarrow $&$   {\rm E}_6$& $\longrightarrow$ &
${\rm SO(10)}$&$ \longrightarrow$ &${\rm SU(5)}$,\cr
&$\epsilon_0$&&$\epsilon_1$&&$\epsilon_2
$
\end{tabular}
\label{breaking}
\end{equation}
which leads to the mass hierarchy
\begin{eqnarray}
m_t \gg m_c \gg m_u \nonumber\\
m_b\gg m_s\gg m_d\\
m_\tau\gg m_\mu\gg m_e.\nonumber
\end{eqnarray}

To realize this hierarchy we assume that
the global E$_7$ is broken explicitly by
the fundamental representation of E$_7, {\bf 56}$,
which contains six breaking parameters,
$\epsilon_0, \bar{\epsilon}_0,\epsilon_1, \bar{\epsilon}_1,
\epsilon_2, \bar{\epsilon}_2 $ that are all singlets of
SU(5). They carry U(1) charges as
\begin{eqnarray}
&\epsilon_0 (-3,0,0), &\bar{\epsilon}_0 (3,0,0)\nonumber\\
&\epsilon_1 (-1,-4,0), &\bar{\epsilon}_1 (1,4,0)\\
&\epsilon_2 (-1,-1,-5), &\bar{\epsilon}_2 (1,1,5)\nonumber
\end{eqnarray}
where the numbers in each parenthesis denote charges
of U(1)$_1\times$U(1)$_2\times$U(1)$_3$. The desired hierarchy
in eq.(\ref{breaking}) is represented by
\begin{equation}
\epsilon_0 \gg \epsilon_1 \gg
\epsilon_2.
\end{equation}

The structure of Yukawa couplings for the NG quarks and leptons
depends on U(1) charges of the Higgs {\bf 5}$_H$ and {\bf 5}$^*_H$.  
To determine them,
we consider that the Higgs multiplets  ${\bf 5}_H$ and
${\bf 5}_H^*$ belong to {\bf 27} of E$_6$ in
{\bf 133} of E$_7$. Then,
U(1) charges for the {\bf 5}$_H$ are given by
\begin{equation}
  {\bf 5}_H \ (2,2,2).
\end{equation}
The Higgs {\bf 5}$^*_H$ is a linear combination of two
{\bf 5}$^*$'s in {\bf 27} of E$_6$\footnote{
{\bf 27} of E$_6$ is decomposed to {\bf 16} + {\bf 10} +
{\bf 1} of SO(10). The {\bf 16} and {\bf 10} contain one {\bf 5} and
two {\bf 5}$^*$ of SU(5).} as
\begin{equation}
  {\bf 5}^*_H = \sin\theta {\bf 5}^*_{16} + \cos\theta {\bf 5}^*_{10}
\end{equation}
where U(1) charges for  {\bf 5}$^*_{16}$ and {\bf 5}$^*_{10}$
are given by\footnote{
The orthogonal combination of the {\bf 5}$_{16}^*$ and {\bf 5}$^*_{10}$
is assumed to have a GUT scale mass. We also assume that color
triplets in {\bf 5}$_H$ and {\bf 5}$^*_H$ receive a GUT scale mass
after the spontaneous breakdown of the SU(5) GUT.
This requires a fine tuning. We do not, however,
discuss this fine tuning problem here, since it is beyond the scope of this
talk.}
\begin{equation}
{\bf 5}^*_{16} (2,-1,3) \ \ {\rm and}
\ \  {\bf 5}^*_{10} (2,2,-2).
\end{equation}

We  now discuss Yukawa couplings for the quark
and lepton multiplets.
In general, Yukawa couplings are given
in a form $a_n\epsilon^n\psi\psi H$ where $\epsilon$, $\psi$
and $H$ stand for the explicit
breaking parameters, the NG multiplets and the Higgs
multiplets, respectively. By our choice of the U(1) charges for the explicit breaking
parameters and Higgs multiplets,
Yukawa couplings take the following form in the leading order of the
explicit breaking
parameters, $\epsilon$'s;
\begin{eqnarray}
  \label{Yukawas}
  W &=& W_U + W_D + W_E + W_\nu,\\
W_U &=& \sum_{ij} a_{ij} Y_{Uij} {\bf 10}_i {\bf 10}_j {\bf 5}_H,
\label{Wu}\\
W_D &=& W_E = \sum_{ij} b_{ij} Y_{D/Eij} {\bf 5}^*_i {\bf 10}_j {\bf 5}^*_H,
\label{Wde}\\
W_\nu &=& \sum_{ij} c_{ij} Y_{\nu ij} {\bf 5}^*_i {\bf 1}_j {\bf 5}_H,
\label{Wnu}
\end{eqnarray}
where $W_U,\ W_D,\ W_E$ and $W_\nu$ represent superpotentials
of Yukawa couplings for
up-type quarks, down-type quarks, charged leptons and neutrinos.
In these expressions $Y$'s are given by,\footnote{
One may wonder that in eq.(\ref{Yde}) the (3,1) element of $Y_{D/E}$,
has a term of $\epsilon_0\epsilon_1$.
We do not think that such a term
appears there, since in the limit $\epsilon_2 \rightarrow 0$, the global
SO(10) symmetry becomes exact and the {\bf 10}$_1$ is the true
NG multiplet which has no Yukawa interaction in the superpotential.
}$^,$
\footnote{Precisely speaking, our coset-space E$_7$/SU(5)$\times$U(1)$^3$
contains three dimensional parameters $f_0,\ f_1$ and $f_2$.
We assume  $f_0\sim f_1\sim f_2$ here, for simplicity. However,
even if it is not the case, one obtains the same form of
Yukawa couplings as in eqs.(\ref{Yu}), (\ref{Yde}) and (\ref{Ynu})
by redefining $\epsilon$'s as $\epsilon_i = \tilde\epsilon_i/f_i$
($i=$0,1,2) where $\tilde\epsilon_i$ are original dimensional parameters
for the explicit E$_7$ breakings.}
\begin{equation}
  \label{Yu}
  Y_U \simeq \left(
\begin{array}{ccc}
\epsilon_2^2&\epsilon_1\epsilon_2&\epsilon_0\epsilon_2\\
\epsilon_1\epsilon_2&\epsilon_1^2&\epsilon_0\epsilon_1\\
\epsilon_0\epsilon_2&\epsilon_0\epsilon_1&\epsilon_0^2\\
\end{array}
\right),
\end{equation}

\begin{equation}
  \label{Yde}
  Y_{D/E} \simeq \left(
\begin{array}{ccc}
\epsilon_1\epsilon_2\cos\theta&\epsilon_1^2\cos\theta
&\epsilon_0\epsilon_1\cos\theta\\
\epsilon_0\epsilon_2\cos\theta&\epsilon_0\epsilon_1\cos\theta
&\epsilon_0^2\cos\theta\\
\epsilon_0\epsilon_2\sin\theta&\epsilon_0\epsilon_1\sin\theta
&\epsilon_0^2\sin\theta\\
\end{array}
\right),
\end{equation}

\begin{equation}
  \label{Ynu}
  Y_\nu \simeq \left(
\begin{array}{ccc}
\epsilon_1^2&\epsilon_0\epsilon_1
&\epsilon_0\epsilon_2\\
\epsilon_0\epsilon_1&\epsilon_0^2&0\\
0&0&\epsilon_0^2
\end{array}
\right)
\end{equation}

We have assumed the E$_7$ representations for $\epsilon_i$, {\bf 5}$_H$
and {\bf 5}$^*_H$ to determine their U(1) charges. However, we
consider that this assumption is over statement since the E$_7$
is already spontaneously broken. What is relevant to our analysis is
only their charges of the unbroken subgroup SU(5)$\times$U(1)$^3$.
With this general consideration it is
impossible to estimate the coefficients 
$a_{ij},\ b_{ij}$ and $c_{ij}$  in eqs.(\ref{Wu}),
(\ref{Wde}) and (\ref{Wnu})
and hence we assume that they are of O(1).

From the above Yukawa couplings in eqs.(\ref{Yu})
and (\ref{Yde}) we easily derive the following mass
relations;
\begin{eqnarray}
&&\displaystyle
\frac{m_u}{m_c} \sim \frac{\epsilon_2^2}{\epsilon_1^2},\nonumber \\ 
&&\displaystyle
\frac{m_c}{m_t} \sim \frac{\epsilon_1^2}{\epsilon_0^2},\nonumber\\
  \label{massrelation}
&&\displaystyle
\frac{m_e}{m_\mu} = \frac{m_d}{m_s} \sim \frac{\epsilon_2}{\epsilon_0}
\sin^{-1}\theta, \\
&&\displaystyle
\frac{m_\mu}{m_\tau} = 
\frac{m_s}{m_b} \sim \frac{\epsilon_1}{\epsilon_0}\sin\theta\cos\theta.
\nonumber
\end{eqnarray}
These relations describe very well the observed mass relations
provided that 
\begin{eqnarray}
  \label{eps}
\frac{\epsilon_1}{\epsilon_0} \sim 0.05,\ \ \frac{\epsilon_2}{\epsilon_1}
\sim 0.05 \ \ {\rm and}\ \ \tan\theta \sim 1.
\end{eqnarray}

We see that the Cabibbo-Kobayashi-Maskawa mixing angles for quarks
between the 1st and the 2nd,
the 2nd and the 3rd, and the 3rd and the 1st family are of the order
$\epsilon_2/\epsilon_1$,
$\epsilon_1/\epsilon_0$, and
$\epsilon_2/\epsilon_0$, respectively. It also describes the observed
mixing angles
 very well provided that the relations in eq.(\ref{eps}) are
satisfied.

We do not further mention details of the mass relations
since there should be corrections to the mass matrices in eqs.(\ref{Yu})
and (\ref{Yde}) from some higher
dimensional operators which
may affect masses for lighter particles significantly.
Otherwise, we have a SU(5) GUT relation, $m_d=m_e$,
which seems unrealistic\cite{PDG}.

So far, we have discussed the mass matrices for quarks and charged
leptons and found that the qualitative global structure of the obtained
matrices fits very well the observed mass spectrum
for quarks and charged leptons (except for $m_d=m_e$)
and mixing angles for quarks
if the relations in eq.(\ref{eps}) are satisfied\footnote{
The observed mass for the strange quark seems somewhat smaller
than the SU(5) GUT value\cite{PDG}.}.

We are now at the point to discuss neutrino masses and lepton mixings.
We assume that  Mayorana masses for right-handed
neutrinos $N_i$ are induced by SU(5) singlet Higgs
multiplets $\bar{s}_i({\bf 1})$.
We introduce two singlets $\bar{s_1}$({\bf 1}) and 
$\bar{s}_2$({\bf 1}) whose U(1) charges\footnote{
These $\bar{s_i}$({\bf 1}) are regarded as SU(5) singlet components
of {\bf 56} of E$_7$.} are given by
\begin{eqnarray}
  \label{singlets}
\bar{s}_1 (1,4,0)\ \ {\rm and}\ \ 
\bar{s}_2 (1,1,5).
\end{eqnarray}
Their vacuum expectation values, $\langle\bar{s_1}\rangle$ and
$\langle\bar{s_2}\rangle$ are expected to be of order
of the SU(5) GUT scale $\sim 10^{16}$ GeV.

Majorana masses for $N_i$ are induced from nonrenormalizable
interactions of a
form;\footnote{Other mass terms such as
$\epsilon^2 N_i N_j$ can be forbidden
by some chiral symmetry.}
\begin{equation}
W_N = \frac{\epsilon^2}{M_G} N_i N_j \bar{s}_k \bar{s}_l.
\end{equation}
Here, $M_G$ is the gravitational scale $M_G\simeq 2.4\times
10^{18}$ GeV.
Then, the matrix of the Majorana masses takes the following
form;\footnote{The mass term of the form $\epsilon^4 N_i N_j$
may produce a similar form to eq.(\ref{MnuR}) if $\bar{\epsilon}_0=0$
and $\bar{\epsilon}_1,\bar{\epsilon}_2 \ne 0$.}
\begin{equation}
  \label{MnuR}
  M_{\nu_R} =  \frac{1}{M_G}\left(
\begin{array}{ccc}
\epsilon_1^2 \bar{s_2}^2
&\epsilon_0\epsilon_1 \bar{s_2}^2&\epsilon_0\epsilon_1 \bar{s_1}
\bar{s_2}\\
\epsilon_0\epsilon_1  \bar{s_2}^2&
\epsilon_0^2 \bar{s_2}^2&\epsilon_0^2  \bar{s_1}\bar{s_2}\\
\epsilon_0\epsilon_1 \bar{s_1}
\bar{s_2}& \epsilon_0^2 \bar{s_1} \bar{s_2}&\epsilon_0^2 \bar{s_1}^2
\end{array}
\right),
\end{equation}
where all elements are multiplied by undetermined factors of O(1) 
like in the case for quarks and leptons.

The neutrino masses are given by\cite{seesaw}
\begin{equation}
  \label{numass}
  m_\nu \simeq m_{\nu_D} M_{\nu_R}^{-1} m_{\nu_D}^T,
\end{equation}
where
\begin{equation}
  \label{nudmass}
   (m_{\nu_D})_{ij} = c_{ij} Y_{\nu ij} \langle{\bf 5}_H \rangle.
\end{equation}
Three eigenvalues of the matrix in eq.(\ref{numass}) are of order,
$m_{\nu_1} \sim \epsilon^2_1 M_G \langle {\bf 5}_H \rangle^2/
\langle \bar{s}_2 \rangle^2,$
$m_{\nu_2} \sim \epsilon^2_0 M_G \langle {\bf 5}_H \rangle^2/
\langle \bar{s}_2 \rangle^2$ and $
m_{\nu_3} \sim \epsilon^2_0 M_G \langle {\bf 5}_H \rangle^2/
\langle \bar{s}_1 \rangle^2$.
It is remarkable that for $\langle {\bf 5}_H \rangle
\sim 100$GeV, $\epsilon_0 \sim 1$ and $\langle \bar{s}_i\rangle
\sim 10^{16}$GeV
we get the desired mass for neutrino $m_{\nu_i} \sim 0.1$ eV.

From the  Mikheev-Smirnov-Wolfenstein solution(MSW)\cite{MSW}
to the solar neutrino problem, we have\cite{BK,nurev}
\begin{equation}
  \delta m_{\nu_e\nu_\mu}^2 \simeq 10^{-6} - 10^{-5} {\rm eV}^2.
\end{equation}
We see that there are two choices
\begin{equation}
  \left(\frac{\langle \bar{s}_1 \rangle}{\langle \bar{s}_2 \rangle}
\right)^2
\sim 10^{-2}-10^{-1}\  {\rm or}\  
  \left(\frac{\langle \bar{s}_2 \rangle}{\langle \bar{s}_1 \rangle}
\right)^2
\sim 10^{-2}-10^{-1}
\end{equation}
to account for atmospheric and solar neutrino anomalies,
simultaneously. Thus, all off-diagonal elements of the 
diagonalization matrix for the neutrino mass matrix in eq.(\ref{numass})
are of O(0.1) in either cases.

However, it is very interesting that 
the mixing angle for lepton doublets
which mixes  charged leptons in the second and the third family
is of order $\tan\theta$ (see eq.(\ref{Yde})) and hence of the order 1.
This means, together with the above result,
that the weak mixing angle relevant for
$\nu_\mu$-$\nu_\tau$ oscillation can be so large,
$\sin^2 2\theta_{\nu_\mu\nu_\tau} \simeq 1$, as required for explaining
the observed atmospheric neutrino anomaly.
On the other hand, the mixing angle for $\nu_\mu-\nu_e$
oscillation is very small, $\theta_{\nu_\mu\nu_e} \sim $ O(0.1),
which may fit the small angle MSW solution\cite{BK,nurev} to the solar
neutrino problem.

\section{Conclusion and Discussion}

In this talk we have shown that the coset-space family unification
on E$_7$/SU(5)$\times$U(1)$^3$ naturally accommodates the large
lepton mixing, $\sin^22\theta_{\nu_\mu\nu_\tau} \simeq 1$,
 necessary for explaining the atmospheric neutrino
anomaly reported by the SuperKamiokande collaboration\cite{SK}.
The main reason why we have a large mixing of the SU(2) lepton doublets
in the second and the third family is the twisted
structure of family. Namely, the {\bf 5}$^*$'s
in the second and the third family both live on the same coset-subspace
E$_7$/E$_6\times$U(1). On the other hand the {\bf 10}'s
in the third, the second and the first family live on the separate 
coset-subspaces, E$_7$/E$_6\times$U(1), E$_6$/SO(10)$\times$U(1)
and SO(10)/SU(5)$\times$U(1), respectively.
This unparallel family structure is an unique feature of
the present coset-space family unification.

It is quite natural that the NG multiplets carry no U(1)$_R$ charge.
Thus, the dangerous lower ($d=4,5$) dimensional operators contributing to
proton decays are forbidden by imposing the R-invariance U(1)$_R$. However,
the R invariance is broken at the gravitino scale at least and hence we may
expect small R-violating $d=4$ operators.

The existence of approximate global E$_7$ symmetry is the most crucial
assumption in our coset-space family unification. We hope that it is
understood by some underlying physics at the gravitational scale. The Horava
and Witten M theory\cite{HW} will be a hopeful example, since
it is known\cite{Sharpe} that there
appear enhanced global symmetries on the 10 dimensional
boundary of 11 dimensional space-time
.

\end{document}